\documentclass[aps,prd,
preprint,
floatfix,nofootinbib,groupedaddress,showpacs]{revtex4}
\usepackage{graphicx}
\usepackage{amsmath}
\usepackage{latexsym}
\usepackage{here}
\usepackage{slashbox}
\usepackage{array}
\usepackage{dcolumn}
\usepackage{bm}
\usepackage{multirow}

\usepackage[usenames,dvipsnames]{color}
\definecolor{Black}{named}{Black}
\definecolor{Red}{named}{Red}
\definecolor{Blue}{named}{Blue}

\renewcommand{\Re}{{\rm Re\thinspace}}

\newcommand{\Lumint}{{\cal L}_{\rm int}}

\def\epem{\ifmmode e^+e^-\else $e^+e^-$\fi}
\def\to{\rightarrow}

\def\mpl{\ifmmode \overline M_{Pl}\else $\bar M_{Pl}$\fi}
\def\beq{\begin{equation}}
\def\be{\begin{equation}}
\def\beqn{\begin{eqnarray}}
\def\ee{\end{equation}}
\def\eeq{\end{equation}}
\def\eeqn{\end{eqnarray}}

\begin{document}

\title{
Unique heavy lepton signature at $e^+e^-$ linear collider with
polarized beams}



\author{
G.
Moortgat-Pick,$^{a,b,}$\footnote{gudrid.moortgat-pick@desy.de}\hspace{.4cm}
P. Osland,$^{c}$\footnote{per.osland@ift.uib.no}\hspace{.4cm} A.
A. Pankov$^{d,}$\footnote{pankov@ictp.it}\hspace{.4cm} A. V.
Tsytrinov$^{d,}$\footnote{tsytrin@rambler.ru}\hspace{.4cm} }
\vspace{.5cm}
\affiliation{ $^{\rm a}$II.\ Inst.\ of Theor.\ Physics, University of Hamburg, 
Luruper Chaussee 149, 22761 Hamburg, Germany\\
$^{\rm b}$DESY, Notkestrasse 85, 22607 Hamburg, Germany\\
$^{\rm c}$Department of Physics and Technology, University of Bergen, Postboks 7803, N-5020  Bergen, Norway\\
 $^{\rm d}$The Abdus Salam ICTP Affiliated Centre, Technical University of Gomel, 246746 Gomel,
 Belarus
 }

\begin{abstract}

We explore the effects of neutrino and electron mixing with exotic
heavy leptons in the process $e^+e^-\to W^+W^-$ within $E_6$
models. We examine the possibility of uniquely distinguishing and identifying such
effects of heavy neutral lepton exchange from $Z$-$Z'$ mixing within
the same class of models and also from analogous ones due to
competitor models with anomalous trilinear gauge couplings (AGC)
that can lead to very similar experimental
signatures at the $e^+e^-$ International Linear Collider (ILC) for 
$\sqrt{s}=350$, 500~GeV and 1~TeV.
Such clear identification of the model is possible by
using a certain double polarization asymmetry. The availability of both
beams being polarized plays a crucial role in identifying such
exotic-lepton admixture. In addition, the sensitivity of the ILC for probing
exotic-lepton  admixture  is substantially enhanced when the
polarization of  the produced $W^\pm$ bosons is considered.
\end{abstract}
\maketitle

\section{Introduction}

Detailed examination of the process
\begin{equation}\label{proc1}
e^+ + e^- \to W^+ + W^-
\end{equation}
at the ILC is a crucial one for studying the electroweak gauge
symmetry, in particular, electroweak symmetry breaking and the
structure of the gauge sector in general, and allows to observe a
manifestation of New Physics (NP) that may appear beyond the
Standard Model (SM). In the SM,  the process (\ref{proc1}) is
described by the amplitudes mediated by photon and $Z$ boson
exchange in the $s$-channel and by neutrino exchange in the
$t$-channel. This reaction is quite sensitive to both the leptonic
vertices and the trilinear couplings to $W^+W^-$ of the SM $Z$ and
of any new heavy neutral boson or a new heavy lepton that can be
exchanged in the $s$-channel or $t$-channel, respectively. A
popular example in this regard, is represented by $E_6$ models
\cite{Langacker:2008yv,Rizzo:2006nw,Leike:1996pj,Leike:1998wr,Riemann:2005es,Hewett:1988xc}.
In particular, an effective $SU(2)_L \times U(1)_Y \times U(1)_{Y'}$
model, which
originates from  the breaking of the exceptional group $E_6$,
leads to extra gauge bosons. Indeed, in the breaking of
this group down to the SM symmetry, two additional neutral gauge bosons could
appear and  the lightest $Z'$  is defined as
\begin{equation}
Z' = Z'_\chi \cos\beta + Z'_\psi \sin \beta
\end{equation}
and can be parametrized in terms of the hypercharges of the two groups
$U(1)_\psi$  and $U(1)_\chi$ 
which are involved in the breaking of the $E_6$ group into a low-energy group of rank 6:
\begin{align}
E_6 &\to SO(10) \times U(1)_\psi \to SU(5) \times U(1)_\chi \times U(1)_\psi \nonumber \\
&\to SU(3)_c \times SU(2)_L \times U(1)_Y \times U(1)_\psi\times U(1)_\chi.
\end{align}
For a sufficiently large  vacuum expectation value of the Higgs field an effective rank-5 model, which leads to the decomposition (see, for example Ref.~\cite{Hesselbach:2001ri})
$SU(3)_c\times SU(2)_L\times U(1)_Y\times U(1)_{Y'}$
can be deduced from the rank-6 model (see below) so that one of the new gauge bosons decouples from low energy phenomenology. The remaining (lighter) new gauge bosons $Z'$ is in general a mixture of $Z_\psi$ and $Z_\chi$ and is assumed to lead to measurable effects at the collider, and 
an angle $\beta$ specifies the
orientation of the $U(1)^{\prime}$ generator in the $E_6$ group space, where
the values $\beta=0$ and $\beta=\pi/2$ would correspond, respectively, to
pure $Z'_\chi$ and $Z'_\psi$ bosons, while the value $\beta=-
\arctan\sqrt{5/3}$ would correspond to a $Z'_\eta$ boson originating from
the direct breaking of $E_6$ to a rank-5 group in superstring
inspired models.

Another characteristic of extended models, apart from the $Z'$, is the
existence of new matter,
new heavy leptons and quarks. 
In $E_6$ models the fermion sector is enlarged, since the matter multiplets
are in larger representations
(the $\underline{27}$ fundamental representation), that contains, in particular, a
vector doublet of leptons.
From the phenomenological point of view it is convenient to classify the
fermions present in $E_6$ in terms of their
transformation properties under $SU(2)$. We denote the particles with
unconventional isospin assignments (right-handed doublets)
as exotic fermions.
We here consider two heavy left- and right-handed $SU(2)$ exotic lepton
doublets \cite{Frampton:1999xi,Langacker:1988ur}
\begin{equation}
\left ( \begin{array}{l}
N\\
E^-
\end{array}\right )_L\;,\;
\left( \begin{array}{l} N\\
E^-
\end{array}\right )_R \;,
\label{doublets}
\end{equation} 
and one $Z'$ boson, with masses larger than
$M_Z$ and coupling constants that may be different from those of the SM. These leptons
are called {\it vector leptons} because both the left- and right-
handed components transform identically under $SU(2)$. We also
assume that the new, ``exotic'' fermions only mix with the
standard ones within the same family (the electron and its
neutrino being the ones relevant to process (\ref{proc1})), which
assures the absence of tree-level generation-changing neutral
currents \cite{Beringer:1900zz}.

Current lower limits on $M_{Z^\prime}$ obtained from  dilepton
pair production at the LHC with $\sqrt{s}=8~\text{TeV}$ and
$\Lumint\approx 20$ fb$^{-1}$ ~\cite{Atlasnote,Chatrchyan:2012it}
 range in the interval $\sim 2.6 - 2.9~\text{TeV}$, depending on the
particular $Z^\prime$ model being tested. 
Already these masses are too high for a $Z^\prime$ to be directly seen at the ILC.
However, even at such high masses,
$Z^\prime$ exchanges can manifest themselves indirectly {\it via}
deviations of cross sections, and in general of the reaction
observables, from the SM predictions.

In this paper, we study the indirect effects induced by heavy
lepton exchange in $W^\pm$
pair production (\ref{proc1}) at the ILC,
with a center of mass energy $\sqrt s=0.5-1~\text{TeV}$ and
time-integrated luminosity of ${\cal L}_{\rm int}= 0.5-1$~ab$^{-1}$. 
We also present results for a  lower energy run at $\sqrt s=350~\text{GeV}$.
For early papers on these effects, see 
Refs.~\cite{Singh:1990rj,Nagawat:1990ui,Babich:1992mu}.
We allow for effects due to extra $Z'$ gauge boson exchange.
Indirect effects may be quite subtle, both when it comes to distinguishing an effect from the SM,
and also as far as the identification
of the source of an observed deviation is concerned, because {\it
a priori} different NP scenarios may lead to the same or similar
experimental signatures. Clearly, then, the discrimination of one
NP model (in our case the $E_6$) from other possible ones needs an
appropriate strategy for analyzing the data.

Recently, the problem of distinguishing the $Z^\prime$ effects,
once observed in process (\ref{proc1}), from the  anomalous gauge
couplings, has been studied in \cite{Andreev:2012cj}. In the AGC
models, there is no new gauge boson exchange, but the $WW\gamma$,
$WWZ$ couplings are modified with respect to the SM values, this
violates the SM gauge cancellation too and leads to deviations of
the cross sections. We consider the CP-conserving set of such couplings, often
referred to as $\kappa_\gamma$, $\kappa_Z$, $\lambda_\gamma$, $\lambda_Z$ 
and $\delta_Z$ \cite{Hagiwara:1986vm,gounaris}.
An alternative effective-field-theory approach to these effects was recently presented \cite{Buchalla:2013wpa}.

In this note, we extend the analysis of Ref.~\cite{Andreev:2012cj},
considering the possibility of uniquely identifying the
effects of heavy neutral lepton exchange from $Z$-$Z'$ mixing within
the same class of $E_6$ models. This is relevant, since in this class of models lepton
mixing and $Z$-$Z^\prime$ mixing can be simultaneously present.
We also distinguish them from analogous ones due to
competitor models with anomalous trilinear gauge couplings in the
process (\ref{proc1}) by exploiting a double polarization asymmetry that
will unambiguously identify  the heavy exotic-lepton mixing
effects\footnote{This approach was recently
exploited for uniquely identifying the indirect effects of
$s$-channel sneutrino exchange against other new physics scenarios
described by contact-like effective interactions in high-energy
$e^+e^-$ annihilation into lepton pairs \cite{Tsytrinov:2012ma}.}
and is only accessible with the
availability of both beams being polarized
\cite{MoortgatPick:2005cw}. 

While the high precision observables determined at LEP severely constrain the electroweak sector \cite{Alcaraz:2007ri}, they leave room for effects at the energies that are discussed here.

The paper is organized as follows. In Section~\ref{sect:lepton-Z-mixing}, we briefly
review the $E_6$ models involving additional $Z^\prime$ bosons and
new heavy charged and neutral leptons and emphasize the role of
the heavy neutral lepton and boson mixings in the process (\ref{proc1}).
Then, in Sect.~\ref{sect:polarized-sigma} we review the structure of the polarized cross section.
In Sect.~\ref{sect:discovery} we determine the discovery reach on the $NWe$ coupling constants,
and in Sect.~\ref{sect:identification} we determine the identification reach, i.e., down to what coupling strength such a heavy neutral lepton can be distinguished from other new-physics effects.
Then, in Sect.~\ref{sect:low-energy} we comment on the 350~GeV option, before concluding in Sect.~\ref{sect:conclusions}.
\section{Lepton and $Z-Z'$ mixing }
\label{sect:lepton-Z-mixing}

\subsection{Weak basis}

To describe  the formalism for mixing among exotic and ordinary
leptons, we start from the leptonic $SU(2)\times U(1)\times
U(1)^{\prime}$ interaction:\footnote{The needed fermion mixing
formalism has been introduced also, e.g., in
\cite{Babich:1992mu}.}
\begin{equation}
-{\cal L}=e\ \left({\tilde J}^{\mu}_\text{em}A_{\mu}
+ {\tilde J}^{\mu}_Z Z_{\mu}
+{\tilde J}^{\mu}_{Z^{\prime}} Z^{\prime}_{\mu}\right)
+ \frac{g}{\sqrt 2}\left({\tilde J}^{\mu}_W W_{\mu}+\text{h.c.}\right),
\label{lagra0}
\end{equation}
where, in the weak-eigenstate basis, and with $V=\gamma,\ Z,\
Z^{\prime}$, the currents in Eq.~(\ref{lagra0}) can be written as:
\begin{equation}
{\tilde J}^{\mu}_V
=\sum_a{\bar\varepsilon}^0_a\gamma^{\mu}Q^{\varepsilon^0}_a
\varepsilon^0_a,\qquad
{\tilde J}^{\mu}_W
=\sum_a{\bar\eta}^0_a\gamma^{\mu}G^{\eta^0}_a \varepsilon^0_a,
\label{currents0}
\end{equation}
where the coupling matrices $Q^{\varepsilon^0}_a$ and $G^{\eta^0}_a$ 
of the neutral and charged currents
are defined by Eqs.~(\ref{const}) and (\ref{charged}) below.
The superscript $``0''$ labels the weak-eigenstate basis.
Furthermore, in Eq.~(\ref{lagra0}) we adopt the following notations:
$e=\sqrt{4\pi\alpha_\text{em}}$, $g=e/s_W$, $s_W=\sin\theta_W$.
 In Eq.~(\ref{currents0}), we
have introduced, with $a=(L,\ R)$ the left- and right-handed
helicities, the charged and neutral leptons by means of the
notation:
\begin{equation}
\varepsilon_a^0 = \left( \begin{array}{c}
e_a^0   \\
E_a^0
\end{array} \right), \qquad
\eta_a^0= \left( \begin{array}{c}
\nu_a^0  \\
N_a^0
\end{array} \right),
\label{zeros}
\end{equation}
where $e$ and $\nu$ are the ordinary SM electron and neutrino, and
$E$ and $N$ are the exotic charged and neutral heavy leptons,
which we assume to be doublets under electroweak $SU(2)$.
Furthermore, the neutral current couplings are represented by the
matrices
$Q^{\varepsilon^0}_a= Q^{\varepsilon^0}_{\text{em},a};\
g_a^{\varepsilon^0};\
g_a^{\prime\varepsilon^0}$, with:
\begin{equation}
{Q^{\varepsilon^0}_{\text{em},a}}
= \left( \begin{array}{cc}
-1 & 0   \\
0 & -1
\end{array} \right),\qquad
{g_a^{\varepsilon^0}}
= \left( \begin{array}{cc}
 g_a^{e^{0}}& 0   \\
0 & g_a^{E^{0}}
\end{array} \right),\qquad
{g_a^{\prime\varepsilon^0}}
= \left( \begin{array}{cc}
 g_a^{{\prime} e^{0}}& 0   \\
0 & g_a^{{\prime}E^{0}}
\end{array} \right),
\label{const}
\end{equation}
for the $\gamma$, $Z$ and $Z^\prime$, respectively,
where ($\varepsilon^0=e^0,\ E^0$)
\begin{equation}
g_a^{\varepsilon^{0}}=(T_{3a}^{\varepsilon^{0}}-Q_{\text{em},a}^{\varepsilon^{0}}s_W^2)
g_Z, \label{sm}
\end{equation}
and $T^{\varepsilon^0}_{3a}$ is the third isospin
component.
Furthermore, $g_Z=1/s_Wc_W$, with $c_W=\cos\theta_W$.

For the $Z^{\prime}$ couplings to fermions in $E_6$ models, we follow the notation of \cite{Babich:1992mu}:
\begin{alignat}{2}
g_L^{{\prime}e^{0}}&=(3A+B) g_{Z'}, &\quad
g_R^{{\prime}e^{0}}&=(A-B) g_{Z'}, \nonumber \\
g_L^{{\prime}E^{0}}&=(-2A-2B) g_{Z'}, &\quad
g_R^{{\prime}E^{0}}&=(-2A+2B)g_{Z'},
\end{alignat}
where $g_{Z^\prime}=1/c_W$, $A=\cos\beta/(2\sqrt 6)$,
$B={\sqrt{10}}\sin\beta/12$. 

The charged current couplings read:
\begin{equation} {G_a^{\eta^{0}}} = \left( \begin{array}{cc}
 G_a^{{\nu}^{0}}& 0   \\
0 & G_a^{{N}^{0}}
\end{array} \right)
\label{charged}
\end{equation}
with $G_L^{\nu^{0}}=1$,  $G_R^{\nu^{0}}=0$,
$G_a^{N^{0}}=-2\,T^E_{3a}$.

\subsection{Fermion mass basis}

We introduce mass eigenstates in the same notation as
(\ref{zeros}):
\begin{equation}
\varepsilon_a = \left( \begin{array}{c}
e_a   \\
E_a
\end{array} \right), \qquad
\eta_a= \left( \begin{array}{c}
\nu_a  \\
N_a
\end{array} \right).
\label{nozeros}
\end{equation}
These states are related to the weak eigenstates (\ref{zeros}) by
the following transformations:
\begin{equation}
\varepsilon_a=U(\psi_{1a})\varepsilon_a^0;\qquad\qquad
\eta_a=U(\psi_{2a})\eta_a^0,\label{transform}
\end{equation} where the
unitary mixing matrices $U(\psi_{1a})$ and $U(\psi_{2a})$
diagonalize, respectively, the charged and neutral
fermion mass matrices. $U(\psi_{1a})$ and $U(\psi_{2a})$ can be written
as:
\begin{equation}
U(\psi_{1a})= \left( \begin{array}{cc}
\cos\psi_{1a}& \sin\psi_{1a}   \\
 -\sin\psi_{1a} & \cos\psi_{1a}
\end{array} \right)\equiv
\left( \begin{array}{cc}
c_{1a}& s_{1a}   \\
 -s_{1a} & c_{1a}
\end{array} \right),
\label{u1}
\end{equation}
\begin{equation}
U(\psi_{2a})= \left( \begin{array}{cc}
\cos\psi_{2a}& \sin\psi_{2a}   \\
 -\sin\psi_{2a} & \cos\psi_{2a}
\end{array} \right)\equiv
\left( \begin{array}{cc}
c_{2a}& s_{2a}   \\
 -s_{2a} & c_{2a}
\end{array} \right).
\label{u2}
\end{equation}
Present limits on $s_{1a}^2$ and $s_{2a}^2$ are in general less than 1-2\% 
\cite{Langacker:1988ur,Nardi:1992am,Nardi:1994iv} and $m_N>100~\text{GeV}$ \cite{Beringer:1900zz}.
In the fermion-mass-eigenstate basis one can rewrite the interaction
Lagrangian (\ref{lagra0}) as:
\begin{equation}-{\cal L}=e\ \left({
J}^{\mu}_\text{em}A_{\mu}+ { J}^{\mu}_Z Z_{\mu}+ {J}^{\mu}_{Z^{\prime}}
Z^{\prime}_{\mu}\right)+ \frac{g}{\sqrt 2}\left({ J}^{\mu}_W
W_{\mu}+\text{h.c.}\right),
\label{lagra}
\end{equation}
where
\begin{equation}
{J}^{\mu}_V=\sum_a{\bar\varepsilon}_a\gamma^{\mu}Q^{\varepsilon}_a
\varepsilon_a,\qquad
{J}^{\mu}_W=\sum_a{\bar\eta}_a\gamma^{\mu}G^{\eta}_a
\varepsilon_a.
\label{currents}
\end{equation}
Since the gauge fields of Eq.~(\ref{lagra}) are the same as those of (\ref{lagra0}), we must have
\begin{equation}
Q^{\varepsilon}_a=U(\psi_{1a})\,Q^{\varepsilon^{0}}_a\,U^{-1}(\psi_{1a}),
\qquad\quad
G_a^{\eta}=U(\psi_{2a})\,G^{\eta^{0}}_a\,U^{-1}(\psi_{1a}),
\label{rotat}\end{equation} and
 $Q^{\varepsilon}_a= Q^{\varepsilon}_{em,a}$,
$g_a^{\varepsilon}$, $g_a^{\prime\varepsilon}$, with
\begin{equation}
{g_a^{\varepsilon}} = \left( \begin{array}{cc}
 g_a^{e}& g_a^{eE}   \\
g_a^{eE}& g_a^{E}
\end{array} \right),\qquad
{g_a^{\prime\varepsilon}} = \left( \begin{array}{cc}
 g_a^{{\prime} e}& g_a^{{\prime} eE}   \\
g_a^{{\prime} eE} & g_a^{{\prime}E}
\end{array} \right),
\qquad {G_a^{\eta}} = \left(
\begin{array}{cc}
 G_a^{{\nu}}&  G_a^{{\nu E}}   \\
G_a^{{N e}} & G_a^{{N}}
\end{array} \right).
\label{consts}
\end{equation}
It is clear that the electromagnetic current remains diagonal
under the rotation (\ref{rotat}), and therefore is not affected by
lepton  mixing.

In the weak charged currents of Eq.~(\ref{currents}) the exotic-lepton mixings modify not  only
the left-handed currents but also induce an admixture with the right-handed currents. 
The off-diagonal term in $J_W^\mu$ of Eqs.~(\ref{currents})--(\ref{consts})
induces $NWe$ couplings which allow an additional $t$-channel
exotic-lepton-exchange contribution for the process (\ref{proc1}) (see Fig.~\ref{Fig:diagrams}).
Parametrization of the mixing-modified $\nu We$ and the mixing-induced $NWe$
couplings are summarized in Eqs.~(\ref{coupl_nu}) and (\ref{coupl_N}), respectively.

From (\ref{rotat}) and (\ref{consts}) one can obtain
expressions for the lepton coupling constants:

\begin{equation}
g_a^e=g_a^{e^0}c_{1a}^2+g_a^{E^0}s_{1a}^2,\qquad\qquad g_a^{\prime
e}=g_a^{\prime e^0}c_{1a}^2+g_a^{\prime E^0}s_{1a}^2;
\label{coupl_e}
\end{equation}
\begin{equation}
G_L^{\nu}=c_{1L}c_{2L}-2\, T_{3L}^E\, s_{1L}\,s_{2L},\qquad\
G_R^{\nu}=-2\, T_{3R}^E\, s_{1R}s_{2R};\label{coupl_nu}
\end{equation}
\begin{equation}
G_L^{N e}=-s_{2L}c_{1L}-2\, T_{3L}^Ec_{2L}\, s_{1L},\qquad G_R^{N
e}=-2\, T_{3R}^E\, c_{2R}s_{1R}. \label{coupl_N}
\end{equation}

\subsection{$Z$-$Z^\prime$ mixing}

Concerning $Z$-$Z^\prime$ mixing, it can be parametrized as
\begin{equation}
\left( \begin{array}{c}
Z_1 \\
Z_2
\end{array} \right) =
\left( \begin{array}{cc}
 \cos\phi & \sin\phi   \\
-\sin\phi & \cos\phi
\end{array} \right)\left( \begin{array}{c}
Z  \\
Z'
\end{array} \right), \label{phi}\end{equation}
where $Z,\ Z^{\prime}$ are weak eigenstates, $Z_1,\ Z_2$ are
mass eigenstates and $\phi$ is the $Z$-$Z^{\prime}$ mixing angle.
Finally, taking Eq.~(\ref{phi}) into account, the lepton neutral
current couplings to $Z_1$ and $Z_2$ are, respectively
\cite{Babich:1992mu}:
\begin{equation} g_{1a}^{e}=g_a^{e}\cos\phi+g_a^{{\prime}e}\sin\phi
\hskip 2pt; \qquad\qquad
g_{2a}^{e}=-g_a^{e}\sin\phi+g_a^{{\prime}e}\cos\phi. \label{gaffi}
\end{equation}
Current limits are of the order $\phi=(2-5)\times10^{-3}$ \cite{Beringer:1900zz}.
\section{Polarized cross section}
\label{sect:polarized-sigma}

\begin{figure}[tbh!] %
\centerline{
\includegraphics{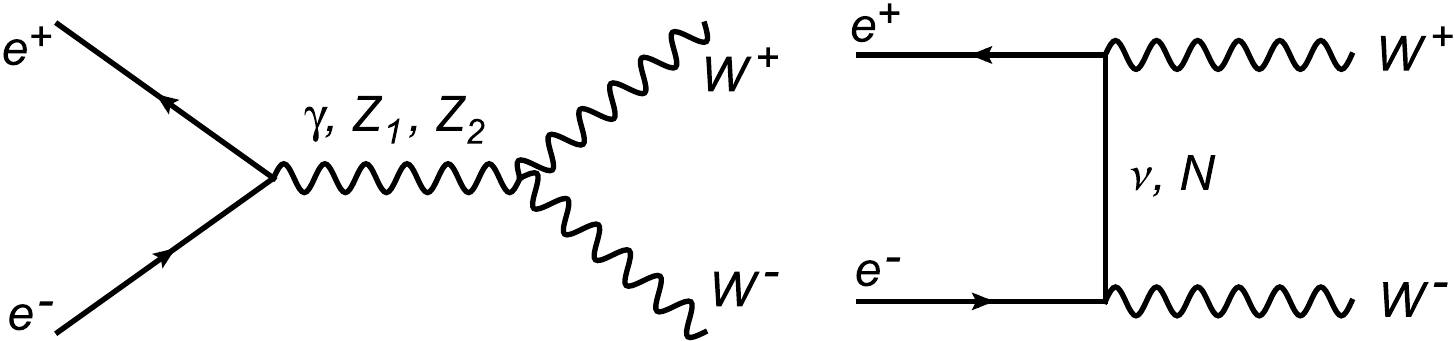}
}
\caption{\label{Fig:diagrams} Feynman diagrams.}
\end{figure}

In the Born approximation the process (\ref{proc1}) is described by the
set of five diagrams shown in Fig.~\ref{Fig:diagrams}  and corresponding to
mass-eigenstate exchanges (i.e. $\gamma$, $\nu$, $N$, $Z_1$ and
$Z_2$), with couplings given by
Eqs.~(\ref{coupl_e})-(\ref{coupl_N}) and (\ref{gaffi}).

The polarized cross section for the process
(\ref{proc1}) can be written  as \cite{Babich:1992mu}
\begin{eqnarray}
\frac{d\sigma\left(P_{L}^-,\hskip 2pt P_{L}^+\right)}{d\cos\theta}
&=&\frac{1}{4}\left[\left(1+P_{L}^-\right)
\left(1-P_{L}^+\right)\hskip 2pt
\frac{d\sigma^{RL}}{d\cos\theta}+\left(1-P_{L}^-\right)
\left(1+P_{L}^+\right)
\hskip 2pt\frac{d\sigma^{LR}}{d\cos\theta}\right.\nonumber \\
&+&\left.\left(1+P_{L}^-\right)\left(1+P_{L}^+\right)\hskip 2pt
\frac{d\sigma^{RR}}
{d\cos\theta}+\left(1-P_{L}^-\right)\left(1-P_{L}^+\right)\hskip 2pt
\frac{d\sigma^{LL}}{d\cos\theta}\right],\label{diffcross}
\end{eqnarray}
where $P_{L}^-$ ($P_{L}^+$) are degrees of longitudinal polarization of
$e^-$ ($e^+$), $\theta$ the scattering angle of the $W^-$ with
respect to the $e^-$ direction.
The superscript ``RL'' refers to a right-handed electron and a left-handed positron, and similarly for the other terms.
The relevant polarized
differential cross sections for $e^-_ae^+_b\to
W^-_{\alpha}W^+_\beta$ contained in Eq.~(\ref{diffcross}) can be
expressed as \cite{Babich:1992mu,Gounaris:1992kp}
\begin{equation}
\frac{d\sigma^{ab}_{\alpha\beta}}{d\cos\theta}= C
\sum_{k=0}^{k=2} F_k^{ab}\hskip 2pt{\cal O}_{k\,\alpha\beta},
\label{A1}
\end{equation}
where $C=\pi\alpha^2_{e.m.}\beta_W/2s$,
$\beta_W=(1-4M_W^2/s)^{1/2}$ the $W$ velocity in the CM frame, and
the helicities of the initial $e^-e^+$ and final $W^-W^+$ states
are labeled as $ab=(RL,\hskip 2pt LR,\hskip 2pt LL,\hskip 2pt RR)$
and $\alpha\beta=(LL,\hskip 2pt TT,\hskip 2pt TL)$, respectively.
The ${\cal O}_k$ are functions of the kinematical variables
dependent on energy $\sqrt{s}$, the scattering  angle $\theta$
and the $W$ mass, $M_W$, which characterize the various
possibilities for the final $W^+W^-$ polarizations ($TT,\hskip 2pt
LL,\hskip 2pt TL+LT$ or the sum over all $W^+W^-$ polarization
states for unpolarized $W$'s).

The $F_k$ are combinations of lepton and trilinear gauge boson
couplings, $g_{WWZ_1}$ and $g_{WWZ_2}$, including lepton and
$Z$-$Z'$ mixing as well as propagators of the intermediate states.
For instance,  for the $LR$ case one finds
\begin{eqnarray}
F_0^{LR} & = & \frac{1}{16s^4_W}\hskip
2pt\left[\left(G_L^{\nu}\right)^2+
r_N \left(G_L^{N e}\right)^2\right]^2, \nonumber \\
F_1^{LR} & = & 2\left[1-g_{WWZ_1}g_{1L}^e\chi_1-
g_{WWZ_2}g_{2L}^e\chi_2\right]^2,\nonumber \\
F_2^{LR} & = & -\frac{1}{2s^2_W}\hskip 2pt
\left[\left(G_L^{\nu}\right)^2+ r_N \left(G_L^{N
e}\right)^2\right]\hskip 2pt \left[1-g_{WWZ_1}g_{1L}^e\chi_1-
g_{WWZ_2}g_{2L}^e\chi_2\right],\label{flr}
\end{eqnarray}
where the $\chi_j$ ($j=1,2$) are the $Z_1$ and $Z_2$ propagators, i.e.
$\chi_j=s/(s-M_j^2+iM_j\Gamma_j)$, $r_N=t/(t-m_N^2)$, with
$t=M_W^2-s/2+s\cos\theta\hskip 1pt\beta_W/2$, and $m_N$ is the
neutral heavy lepton mass. Also, in Eq.~(\ref{flr}),
$g_{WWZ_1}=g_{WWZ}\cos\phi$ and $g_{WWZ_2}=-g_{WWZ}\sin\phi$ where
$g_{WWZ}=\cot\theta_W$. Note that Eq.~(\ref{flr}) is obtained in
the approximation where the imaginary parts of the $Z_1$ and $Z_2$
boson propagators are neglected, which is fully appropriate far away from the poles.
(Accounting for this  effect would
require the replacements $\chi_j\to \Re\chi_j$ and
$\chi_j^2\to\vert\chi_j\vert^2$ on the right-hand side
of Eq.~(\ref{flr}).)

Since the gauge eigenstate $Z'$ is neutral under $SU(2)_L$ and does not
couple to the $W^+W^-$ pair, the process
(\ref{proc1}) is sensitive to a $Z^\prime$ only in the case of a
non-zero $Z$-$Z'$ mixing. Moreover, as one can easily see from the
formulae above, the $s$-channel $Z_2$ and the $t$-channel $N$
exchange amplitudes arise only in the case of non-vanishing mixing
angles. In this case, the expression for the SM cross section
\cite{Gounaris:1992kp} can be obtained from (\ref{diffcross}) in
the limit of vanishing mixing angles.

The first term $F_0^{LR}$ describes the contributions to the cross
section caused by neutrino $\nu$ and heavy neutral lepton $N$
exchanges in the $t$-channel while the second one, $F_1^{LR}$, is
responsible for $s$-channel exchange of the photon $\gamma$ and the gauge
bosons $Z_1$ and $Z_2$. The interference between $s$- and
$t$-channel amplitudes is contained in the term $F_2^{LR}$. The
$RL$ case is simply obtained from Eq.~(\ref{flr}) by exchanging
$L\to R$.

For the $LL$ and $RR$ cases there is only $N$-exchange contribution,
\begin{eqnarray}
F_0^{LL} = F_0^{RR} = \frac{1}{16s^4_W}\hskip 2pt r_N^2 \hskip 2pt
\left(G_L^{N e} G_R^{N e}\right)^2. \label{fll}
\end{eqnarray}
Concerning the ${\cal O}_{k\,\alpha\beta}$ multiplying the expression in Eq.~(\ref{fll})
(see Eq.~(\ref{A1}))
their explicit expressions for polarized and unpolarized final
states $W^+W^-$ can be found in, e.g. \cite{Babich:1992mu}.

\section{Discovery reach on  heavy lepton couplings}
\label{sect:discovery}

We take ``discovery'' of new physics to mean exclusion of the Standard Model at a given confidence level. 
In the following, this will be the 95\% C.L.

\subsection{No $Z$-$Z^\prime$ mixing}

Let us start the analysis with a case where there is only lepton
mixing and no $Z$-$Z'$ mixing, i.e., $\phi=0$. Since the mixing
angles are bounded by $s_i^2$ at most of order $10^{-2}$,
we can expect that retaining only the terms of order
$s_1^2$, $s_2^2$ and $s_1s_2$ in the cross section
(\ref{diffcross}) should be an adequate approximation. To do that
 we  expand the  couplings of Eqs.~(\ref{coupl_e})-(\ref{coupl_N})
 taking Eq.~(\ref{sm}) into account. We find for $E_6$ models, where $T_{3L}^E=T_{3R}^E=-1/2$:
\begin{eqnarray}
&G_L^{Ne}=s_{1L}-s_{2L}, \qquad\qquad &G_R^{Ne}=s_{1R} \nonumber \\
&g_L^e=g_L^{e^0},
&g_R^e=g_R^{e^0}-\frac{1}{2}(G_R^{N e})^2g_Z,\nonumber\\
&G_L^{\nu}=G_L^{\nu^0}-\frac{1}{2}(G_L^{N e})^2, \qquad
&G_R^{\nu}=s_{1R}\hskip 2pt s_{2R}. \label{approx}
\end{eqnarray}
From Eqs.~(\ref{flr})-(\ref{approx}) one can see that in the
adopted approximation the cross section (\ref{diffcross}) allows
to constrain basically the pair of heavy lepton couplings squared,
$((G_L^{N e})^2, (G_R^{N e})^2)$, it is not possible to constrain
$s_{2R}^2$, which represents mixing in the right-handed neutral-lepton sector.

The sensitivity of the polarized differential cross section
(\ref{diffcross}) to the couplings $G_L^{N e}$ and $G_R^{N e}$ is
evaluated numerically by dividing the angular range
$\vert\cos\theta\vert\leq 0.98$ into 10 equal bins, and defining a
$\chi^2$ function in terms of the expected number of events $N(i)$
in each bin for a given combination of beam polarizations
\cite{Andreev:2012cj}:
\begin{equation}
\chi^{2}=
\sum_{\{P_L^-,\,{P}_L^+\}}\sum^\text{bins}_i\left[\frac{N_{\text{SM+NP}}(i)-N_\text{SM}(i)}
{\delta N_\text{SM}(i)}\right]^2,\label{chi2}\end{equation} where
$N(i)=\Lumint\,\sigma_i\,\varepsilon_W$ with $\Lumint$ the
time-integrated luminosity. Furthermore,
\begin{equation}
\sigma_i=\sigma(z_i,z_{i+1})= \int
\limits_{z_i}^{z_{i+1}}\left(\frac{d\sigma}{dz}\right)dz,
\label{sigmai}
\end{equation}
where $z=\cos\theta$ and polarization indices have been
suppressed. Also, $\varepsilon_W$ is the efficiency for $W^+W^-$
reconstruction, for which we take the channel of lepton pairs
($e\nu+\mu\nu$) plus two hadronic jets, giving
$\varepsilon_W\simeq 0.3$ basically from the relevant branching
ratios.  The  procedure outlined above is followed to evaluate
both $N_\text{SM}(i)$ and $N_{\text{SM+NP}}(i)$.

The uncertainty on the number of events $\delta N_\text{SM}(i)$
combines both statistical and systematic errors where the
statistical component is determined by $\delta
N^\text{stat}_\text{SM}(i)= \sqrt{N_\text{SM}(i)}$. Concerning
systematic uncertainties, an important source is represented by
the uncertainty on beam polarizations, for which we assume $\delta
P_L^-/P_L^-= \delta { P_L^+}/{ P_L^+}= 0.5\%$ with the
``standard'' envisaged values $\vert P_L^-\vert=0.8$ and $\vert
{P}_L^+\vert=0.6$ \cite{MoortgatPick:2005cw}. As for the
time-integrated luminosity, for simplicity we assume it to be
equally distributed between the different polarization
configurations. Another source of systematic uncertainty
originates from the efficiency of reconstruction of $W^\pm$ pairs
which we assume to be $\delta\varepsilon_W/\varepsilon_W=0.5\%$.
Also, in our numerical analysis to evaluate the sensitivity of the
differential distribution to model parameters we include
initial-state QED corrections to on-shell $W^\pm$ pair production
in the flux function approach
\cite{Beenakker:1991jk,Beenakker:1990sf,Beenakker:1994vn,Denner:2005es,Denner:2005fg}
that assures a good approximation within the expected accuracy of
the data.

As a criterion to derive constraints on the coupling constants
in the case where no deviations from the SM were observed within
the foreseeable uncertainties on the measurable cross sections, we
impose that
\begin{equation} \label{Eq:chi_sq}
\chi^2\leq \chi^2_{\mathrm{min}} + \chi^2_\text{CL},
\end{equation}
where $\chi^2_\text{CL}$ is a number that specifies the chosen
confidence level, and $\chi^2_{\mathrm{min}}$ is the minimal value of
the $\chi^2$ function.
\begin{figure}[h t p b]
\includegraphics[scale=0.8]{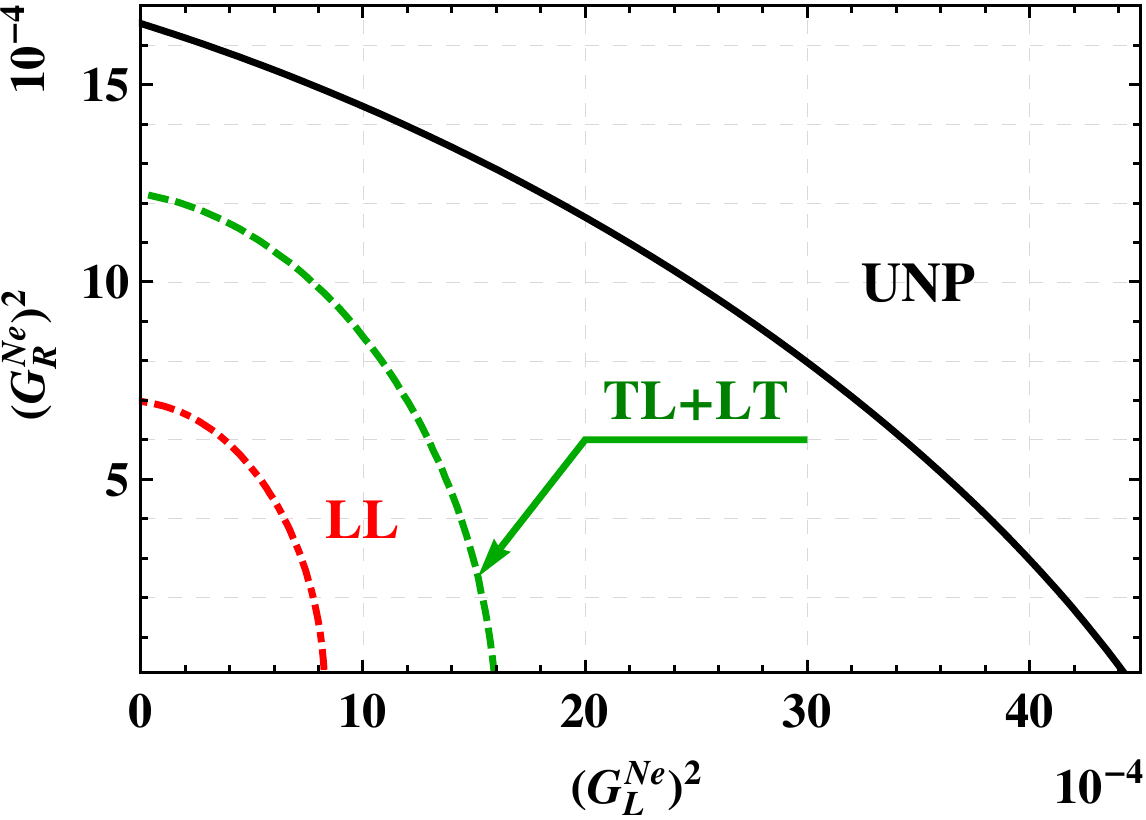}
\caption{Discovery reach (95\% C.L.) on the heavy neutral lepton
couplings $(G_L^{N e})^2$ and $(G_R^{N e})^2$ obtained from
differential polarized cross sections with ($P_{L}^-=\pm
0.8,\;{P}_{L}^+=\mp 0.6$) and different sets of $W^\pm$
polarizations.
 Here,
$\sqrt{s}=0.5~\text{TeV}$,  $\Lumint=0.5~\text{ab}^{-1}$ and
$m_N=0.3$ TeV.}
\label{fig2}
\end{figure}

From the numerical procedure outlined above, we obtain the allowed
regions in $(G_L^{N e})^2$ and $(G_R^{N e})^2$ determined from the
differential polarized cross sections with different sets of
polarization (as well as from the unpolarized process
(\ref{proc1})) depicted in Fig.~\ref{fig2}, where
$\Lumint=500~\text{fb}^{-1}$ has been taken
\cite{MoortgatPick:2005cw}. 

The results of a further potential extension of the present analysis are
also shown in Fig.~\ref{fig2} where the feasibility of measuring
polarized $W^\pm$ states in the process (\ref{proc1}) is assumed.
This assumption is based on the experience gained at LEP2 on
measurements of $W$ polarisation \cite{LEP2Wpolar}. The method exploited
for the measurement of $W$ polarisation is based on the spin density
matrix elements that allow to obtain the differential cross sections for
polarised
$W$ bosons. Information on spin density matrix elements as
functions of the $W^-$ production angle with respect to the
electron beam direction was extracted from the decay angles of the
charged lepton in the $W^-$ ($W^+$) rest frame.
The relevant
theoretical framework for measurement of $W^\pm$ polarisation was
described in \cite{gounaris,Gounaris:1992kp}. 

In Fig.~\ref{fig2}, we consider different cases of polarized $W$s, with $W_L$ and $W_T$ referring to longitudinally and transversely polarized $W$s, respectively.
As shown in the figure,
${d\sigma(W^+_LW^-_L)}/{dz}$ is most sensitive to the parameters
$(G_L^{N e})^2$ and $(G_R^{N e})^2$ while
${d\sigma(W^+_TW^-_T)}/{dz}$ has the lowest sensitivity to those
parameters. The bounds on heavy lepton couplings obtained
from ${d\sigma(W^+_TW^-_T)}/{dz}$ are not presented here as they
are outside of the range shown in
Fig.~\ref{fig2}. The role of $W$ polarization is seen to be
essential in order to set meaningful finite bounds on the $NWe$
couplings.

The obtained bounds are reminiscent of arcs of circles in the $(G_L^{N e})^2$-$(G_R^{N e})^2$ plane. This reflects the fact that the deviations in the $LR$ and $RL$ cross sections are approximately the same for the right-handed and left-handed couplings (recall that $T_{3L}^E=T_{3R}^E$) and thus approximately behave as $(G_L^{N e})^4+(G_R^{N e})^4$.

\begin{figure}[h t p b]
\includegraphics[scale=0.7]{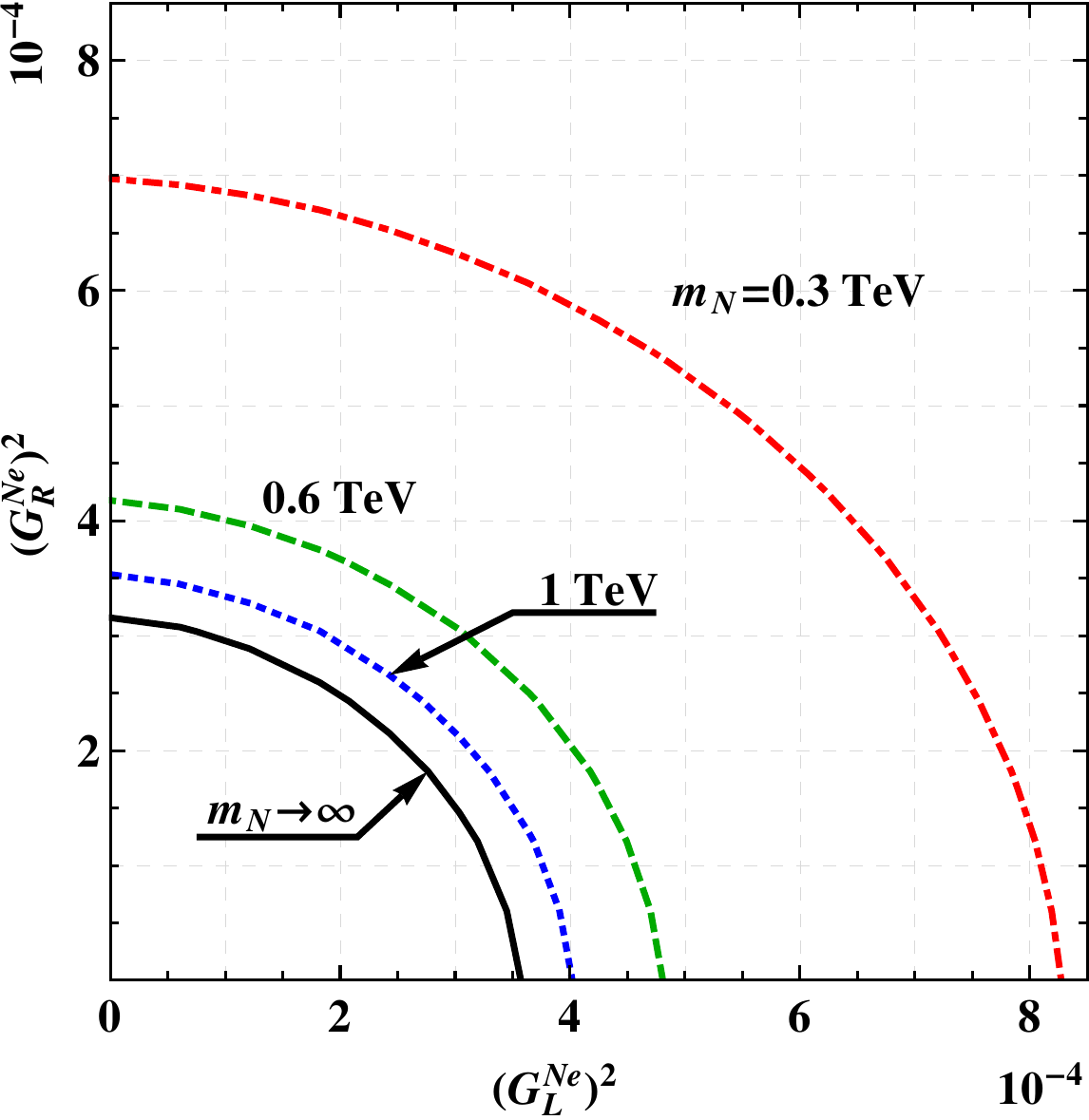}
\caption{Same as in Fig.~\ref{fig2} but obtained from the differential
polarized cross sections ${d\sigma(W^+_LW^-_L)}/{dz}$ only, with
($P_{L}^-=\pm 0.8,\;{P}_{L}^+=\mp 0.6$) and different values of
the lepton mass $m_N=0.3$ TeV, 0.6 TeV, 1 TeV and $m_N\to \infty$.
 Here,
$\sqrt{s}=0.5~\text{TeV}$ and $\Lumint=0.5~\text{ab}^{-1}$.}
\label{fig3}
\end{figure}

In this Fig.~\ref{fig2}, we considered a fairly low mass, $m_N=0.3~\text{TeV}$.
As one can see from Fig.~\ref{fig3}  the constraints on heavy
lepton couplings  become more severe for larger values of $m_N$. The point
is that the deviation of the cross section induced by the lepton
mixing, from the SM prediction can be expressed, e.g., for the LR case,
as
\begin{equation}
\Delta\sigma_{LR}\equiv \sigma_{\rm NP}- \sigma_{\rm
SM}\propto (G_L^{N e})^2\, (1-r_N), 
\end{equation}
where we have used Eqs.~(\ref{flr}) and (\ref{approx}).
This structure $(1-r_N)$ arises from negative interference between a mixing contribution to $\nu$ exchange and the $N$-exchange contribution.
It reflects the
decreasing impact of the heavy neutrino exchange contribution
to $\Delta\sigma_{LR}$, since at large values of $m_N$ the last term will be small. This leads to a
better sensitivity on the mixing angles with increasing $m_N$. The
analogous dependence also holds for $\Delta\sigma_{RL}$ case.

\begin{figure}[tbh!] 
\centerline{
\includegraphics[width=0.49\textwidth]{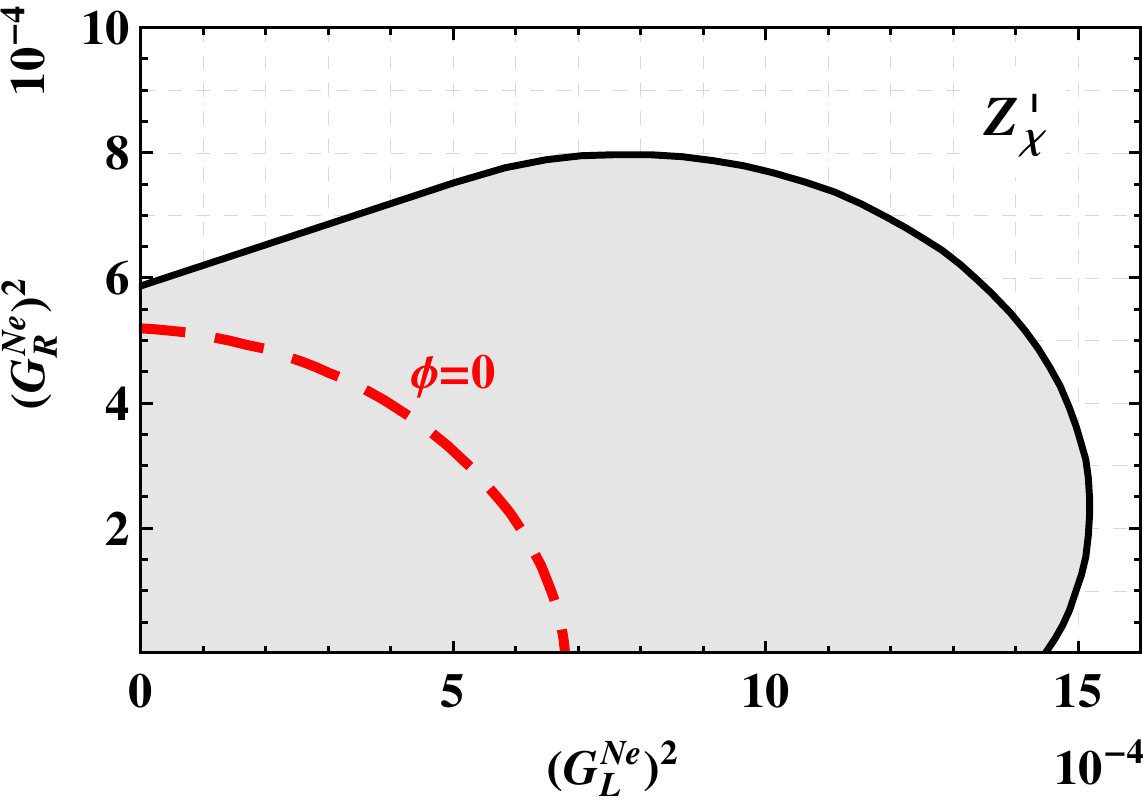}
\includegraphics[width=0.49\textwidth]{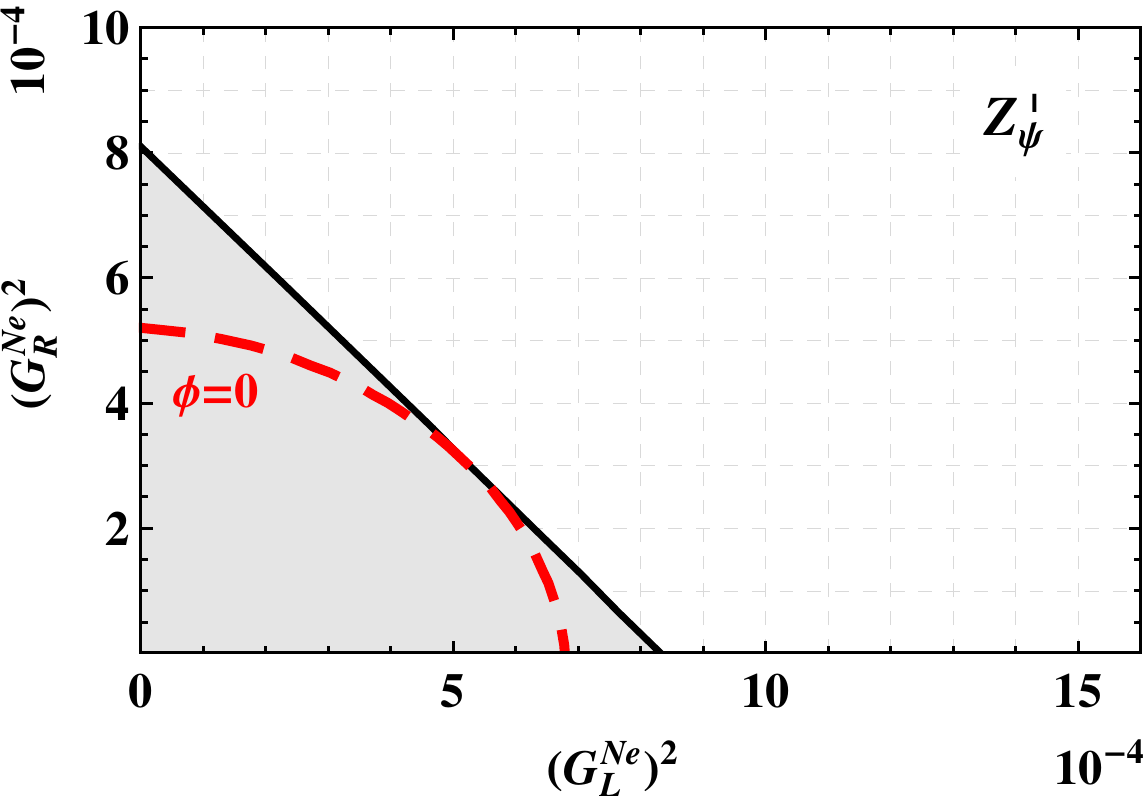}
}
\caption{\label{fig6} Discovery reach at 95\% CL on the heavy
neutral lepton coupling plane $((G_L^{N e})^2,\, (G_R^{N e})^2)$
at $m_N=0.3$ TeV in the case where both lepton mixing and $Z$-$Z^\prime$
mixing are simultaneously allowed for the $Z_\chi^\prime$ model (left
panel) and the $Z_\psi^\prime$ model (right panel), obtained from combined
analysis of polarized differential cross sections
${d\sigma(W^+_LW^-_L)}/{dz}$ at different sets of polarization,
$P_{L}^-=\pm 0.8,\;{P}_{L}^+=\mp 0.6$, at the ILC with
$\sqrt{s}=0.5$ TeV and $\Lumint=1~\text{ab}^{-1}$. 
The dashed curves labelled ``$\phi=0$'' refer to the case of no $Z$-$Z^\prime$ mixing.}
\end{figure}

\subsection{Including $Z$-$Z^\prime$ mixing}

Now we turn to the generic case where both lepton mixing and
$Z$-$Z^\prime$ mixing occur, so that the leptonic coupling constants
are as in Eq.~(\ref{gaffi}) and the $Z_1,\hskip 2pt Z_2$ couplings
to $W^\pm$ are as in Eq.~(\ref{flr}). 
In this case, in order to
evaluate the influence of the $Z$-$Z'$ mixing on the allowed
discovery region on the heavy lepton coupling plane $((G_L^{Ne})^2,\, (G_R^{N e})^2)$ one should vary the mixing angle $\phi$
within its current constraints which depend on the specific $Z'$
model \cite{Erler:2009dq}, namely $-0.0018<\phi<0.0009$ for the $\psi$
model and $-0.0016<\phi<0.0006$ for the $\chi$ model. Within a
specific $Z'$ model and with fixed $m_N$, the $\chi^2$ function
basically depends on three parameters: $\phi$, ${G_L^{N e}}$ and
${G_R^{N e}}$. In this case, Eq.~(\ref{Eq:chi_sq}) describes a
tree-dimensional surface. Its projection on the $((G_L^{N e})^2,\,
(G_R^{N e})^2)$ plane demonstrates the interplay between leptonic
and $Z$-$Z'$ mixings. Fig.~\ref{fig6} shows, as a typical example, the
results of this analysis for the $\chi$-model (left panel) and the
$\psi$-model (right panel), respectively, with fixed $m_N=0.3$
TeV. As one can see, the shapes of the allowed regions for the
coupling constants ${G_L^{N e}}$ and ${G_R^{N e}}$ are quite dependent on the $Z'$
model and different for these two cases. From the
explicit calculation it turns out that this is due to the
different relative signs between the lepton and $Z$-$Z^\prime$ mixing
contributions to the deviations of the cross section
$\Delta\sigma$.

Concerning Fig.~\ref{fig6} and the corresponding analysis for the
$\chi$ and $\psi$ models, we should note that the bounds on the
lepton couplings $(G_L^{N e})^2$ and $(G_R^{N e})^2$ are somewhat
looser than in the case $\phi=0$ discussed above (roughly, by a
factor as large as two), but still numerically competitive with
the current situation.  Also, we can remark that the cross
sections for longitudinal $W^+W^-$ production provide by
themselves the most stringent constraints for this model.

Finally, one should note that although the discovery reach on the lepton
couplings $(G_L^{N e})^2$ and $(G_R^{N e})^2$ obtained from
polarized differential cross sections is quite dependent on the $Z'$
model, this is not the case for the identification reach as the double
beam polarization asymmetry $A_{\rm double}^N$ is basically
independent of the $Z$-$Z'$ boson mixing.

\section{Identification of heavy lepton effects with $A_{\rm double}$}
\label{sect:identification}

By ``identification'' we shall here mean {\it exclusion} of a certain set of competitive models, including the SM, to a certain confidence level. For this purpose,
the double beam polarization asymmetry, defined as
\cite{Rizzo:1998vf,Osland:2003fn,Tsytrinov:2012ma}
\begin{equation}
A_{\rm double}=\frac
{\sigma(P_1,-P_2)+\sigma(-P_1,P_2)-\sigma(P_1,P_2)-\sigma(-P_1,-P_2)}
{\sigma(P_1,-P_2)+\sigma(-P_1,P_2)+\sigma(P_1,P_2)+\sigma(-P_1,-P_2)},
\label{double}
\end{equation}
is very useful.
Here $P_1=\vert{P_{L}^-}\vert$, $P_2=\vert{P_{L}^+}\vert$, and
$\sigma(\pm P_1,\pm P_2)$ denotes the polarized integrated cross
section determined within the allowed range of the $W^-$ scattering angle
(or $\cos\theta$). From Eqs.~(\ref{diffcross}) and (\ref{double})
one finds for the $A_{\rm double}$ of the process (\ref{proc1})
\begin{equation}
A_{\rm
double}=P_1P_2\,\frac{(\sigma^{RL}+\sigma^{LR})-(\sigma^{RR}+\sigma^{LL})}
{(\sigma^{RL}+\sigma^{LR})+(\sigma^{RR}+\sigma^{LL})}.
\label{adouble}
\end{equation}
We note that this asymmetry is only available if both initial beams are polarized.

\begin{figure}[tbh!] 
\centerline{
\includegraphics[width=0.49\textwidth]{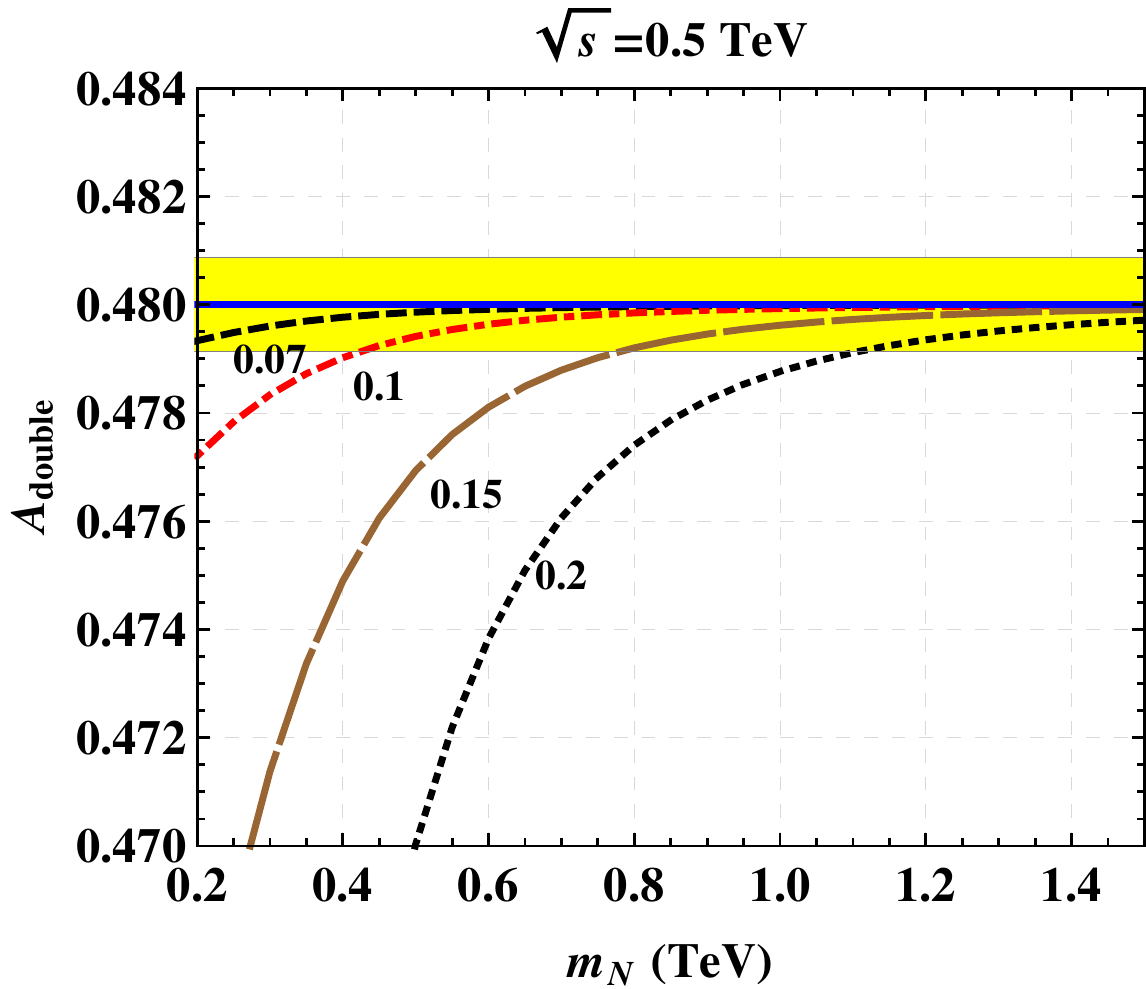}
\includegraphics[width=0.49\textwidth]{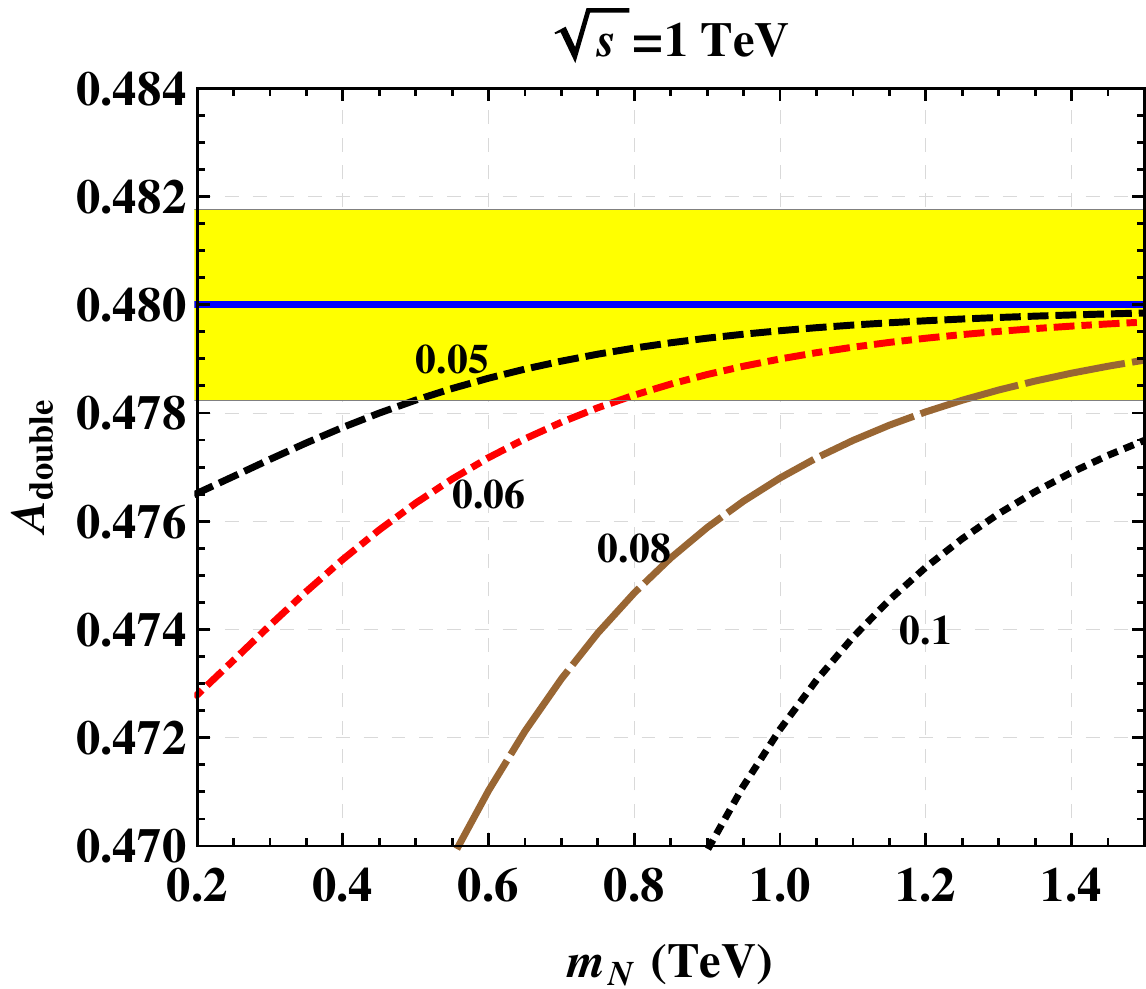}
}
\caption{\label{fig4} Double beam polarization asymmetry $A_{\rm
double}$ for the production of unpolarized $W^\pm$ as a function of
neutral heavy lepton mass $m_N$ for different choices of couplings
$\sqrt{G_L^{N e} G_R^{N e}}$ (attached to the lines) at the ILC
with $\sqrt{s}=0.5$ TeV (left panel) and $\sqrt{s}=1.0$ TeV (right
panel), $\Lumint=1~\text{ab}^{-1}$. The solid  horizontal line
corresponds to $A_{\rm double}^{\rm SM}= A_{\rm double}^{\rm
Z'}=A_{\rm double}^{\rm AGC}$. The error bands indicate the
expected uncertainty in the SM case at the 1-$\sigma$~level.}
\end{figure}

It is important to also note that the SM  gives rise only to
$\sigma^{LR}$ and $\sigma^{RL}$ such that the structure of the
integrated cross section has the form
\begin{equation}
{\sigma_{\rm SM}} =\frac{1}{4}\left[\left(1+P_{L}^-\right)
\left(1-P_{L}^+\right)\hskip 2pt {\sigma^{RL}_{\rm
SM}}+\left(1-P_{L}^-\right) \left(1+P_{L}^+\right) \hskip 2pt
{\sigma^{LR}_{\rm SM}}\right].\label{diffcross_SM}
\end{equation}
This is also the case for anomalous gauge couplings (AGC) \cite{Gounaris:1992kp}, and
$Z'$-boson exchange (including $Z$-$Z'$ mixing and $Z_2$ exchange)
\cite{Andreev:2012cj}. The corresponding expressions for those
cross sections can be obtained from (\ref{diffcross_SM}) by
replacing the specification SM $\to$ AGC and $Z'$,
respectively. Accordingly, the double beam polarization asymmetry has
a common form for all those cases:
\begin{equation}
A_{\rm double}^{\rm SM}=A_{\rm double}^{\rm AGC}= A_{\rm
double}^{\rm Z'}=P_1P_2=0.48, \label{ASM}
\end{equation}
where the numerical value corresponds to the product of the electron and positron
degrees of polarization: $P_1=0.8$, $P_2=0.6$. Eq.~(\ref{ASM})
demonstrates that $A_{\rm double}^{\rm SM}$,  $A_{\rm double}^{\rm
AGC}$ and $A_{\rm double}^{\rm Z'}$ are indistinguishable for any
values of NP parameters, AGC or $Z'$ mass and strength of $Z$-$Z'$
mixing, i.e. $\Delta A_{\rm double}= A_{\rm double}^{\rm
AGC}-A_{\rm double}^{\rm SM}=A_{\rm double}^{\rm Z'}-A_{\rm
double}^{\rm SM}=0$.

On the contrary, the heavy neutral lepton $N$-exchange in the
$t$-channel will induce non-vanishing contributions to $\sigma^{LL}$ and $\sigma^{RR}$, and thus
force $A_{\rm double}$ to a smaller value, $\Delta
A_{\rm double}= A_{\rm double}^{N}-A_{\rm double}^{\rm SM} \propto
-P_1P_2\, r_N^2 \, \left(G_L^{N e} G_R^{N e}\right)^2<0$
irrespectively of the simultaneous lepton and $Z$-$Z'$ mixing
contributions to $\sigma^{RL}$ and $\sigma^{LR}$. A value of
$A_{\rm double}$ below $P_1P_2$ can provide a signature of heavy
neutral lepton $N$-exchange  in the process (\ref{proc1}). All
those features in the $A_{\rm double}$ behavior are shown in
Fig.~\ref{fig4}, where we consider unpolarized $W$s.

\begin{figure}[tbh!] %
\centerline{
\includegraphics[width=0.45\textwidth]{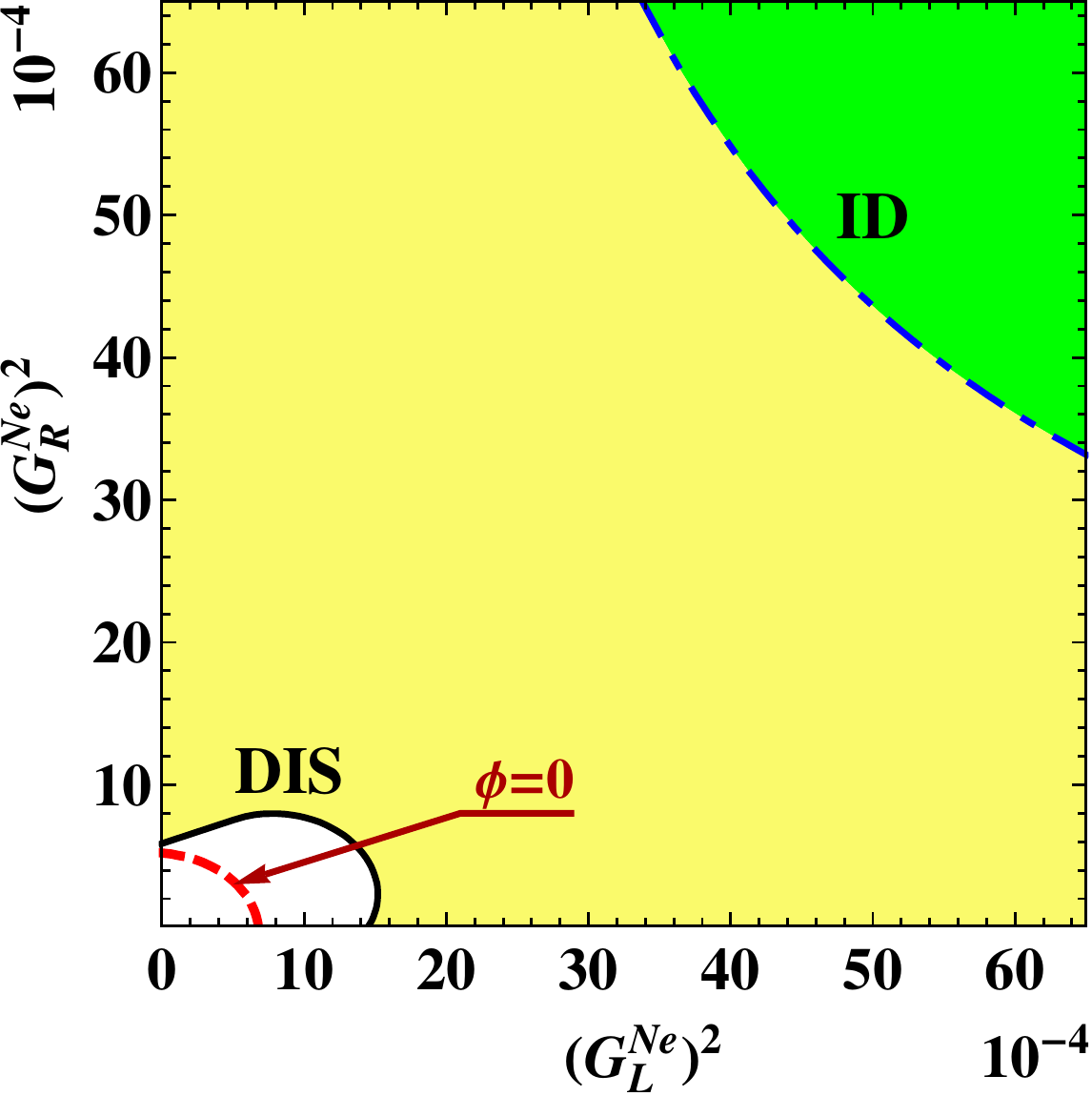}
\includegraphics[width=0.45\textwidth]{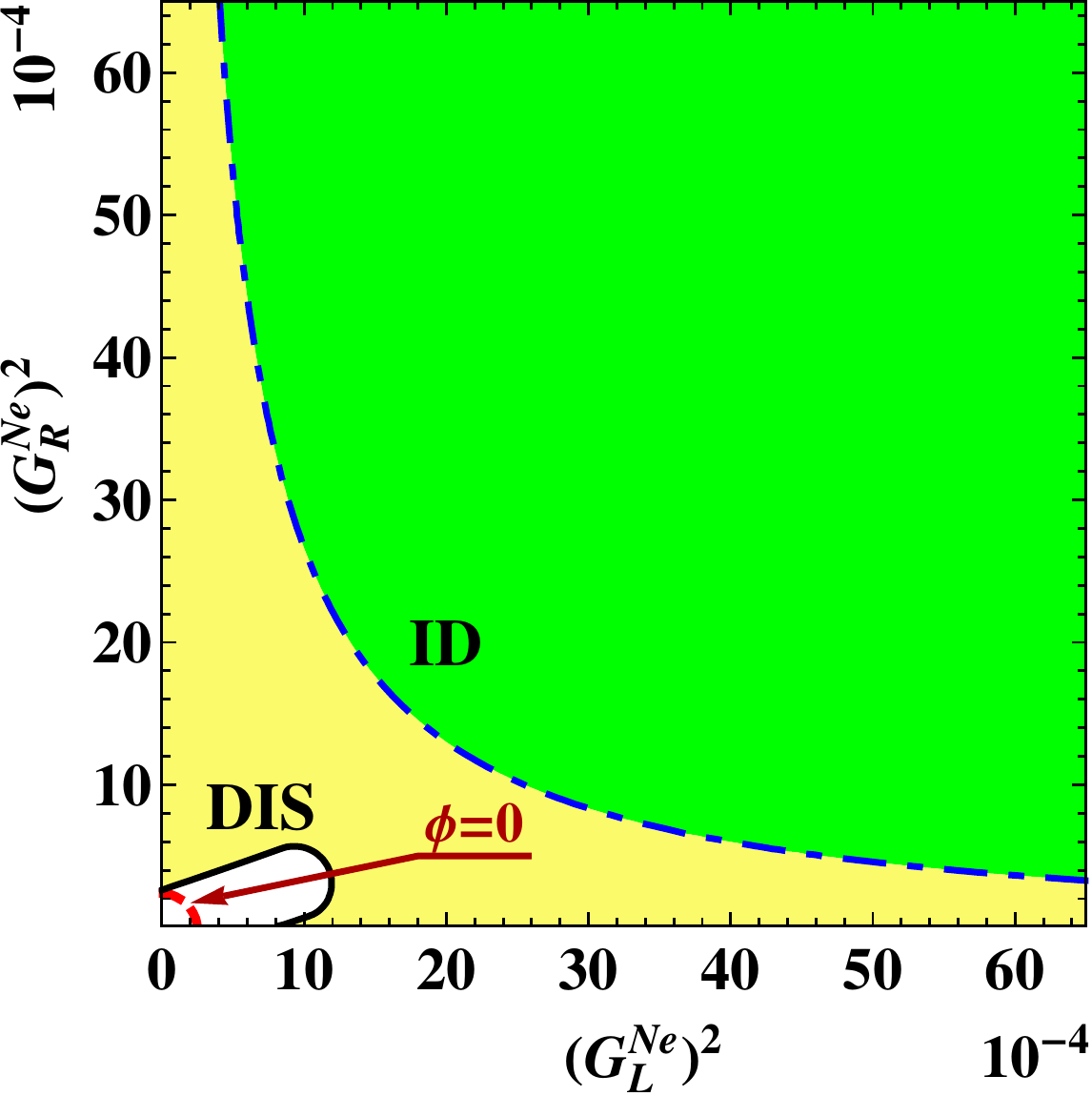}
}
\caption{\label{fig5}  Left panel: discovery (DIS) and
identification (ID) reaches at 95\% CL on the heavy neutral lepton
coupling plane $((G_L^{N e})^2,\, (G_R^{N e})^2)$,
obtained from a combined analysis of polarized differential cross
sections ${d\sigma(W^+_LW^-_L)}/{dz}$ at different sets of
polarization, $P_{L}^-=\pm 0.8,\;{P}_{L}^+=\mp 0.6$, and exploiting the
double polarization asymmetry.
Furthermore, $m_N=0.3$~TeV,
$\sqrt{s}=0.5$ TeV and $\Lumint=1~\text{ab}^{-1}$.  Right panel:
similar, with $\sqrt{s}=1.0$ TeV and for $m_N=0.6$~TeV.
The dashed curves labelled ``$\phi=0$'' refer to the case of no $Z$-$Z^\prime$ mixing,
whereas the outer contour labelled ``DIS'' refer to the minimum discovery reach in the presence of mixing.}
\end{figure}

The identification reach (ID) on the plane of heavy lepton coupling
$((G_L^{N e})^2,\, (G_R^{N e})^2)$ (at 95\% C.L.) for various lepton masses $m_N$
plotted in Fig.~\ref{fig5} is obtained from
conventional $\chi^2$ analysis with $A_{\rm double}$. In that case
the $\chi^2$ function is constructed as $\chi^2=(\Delta
A_{\rm double}/\delta A_{\rm double})^2$ where $\delta A_{\rm
 double}$ is the expected experimental uncertainty accounting for both statistical and
 systematic components. Note that discovery is possible in the green and yellow regions, whereas identification is only possible in the green region.
 The hyperbola-like limit of the identification reach is due to the appearance of a product of the squared couplings
 $(G_L^{N e})^2$ and $(G_R^{N e})^2$ in the deviation from the SM cross section, given by Eq.~(\ref{fll}).

It should be stressed that the identification reach is independent of the $Z^\prime$ model assumed, whereas the discovery reach is not. In fact, in the lower left corner of these figures, we show how the discovery reach gets modified if we allow for $Z$-$Z^\prime$ mixing within the $Z^\prime_\chi$ model (cf.\ Fig.~\ref{fig6}).

\begin{figure}[tbh!] 
\centerline{
\includegraphics[width=0.7\textwidth]{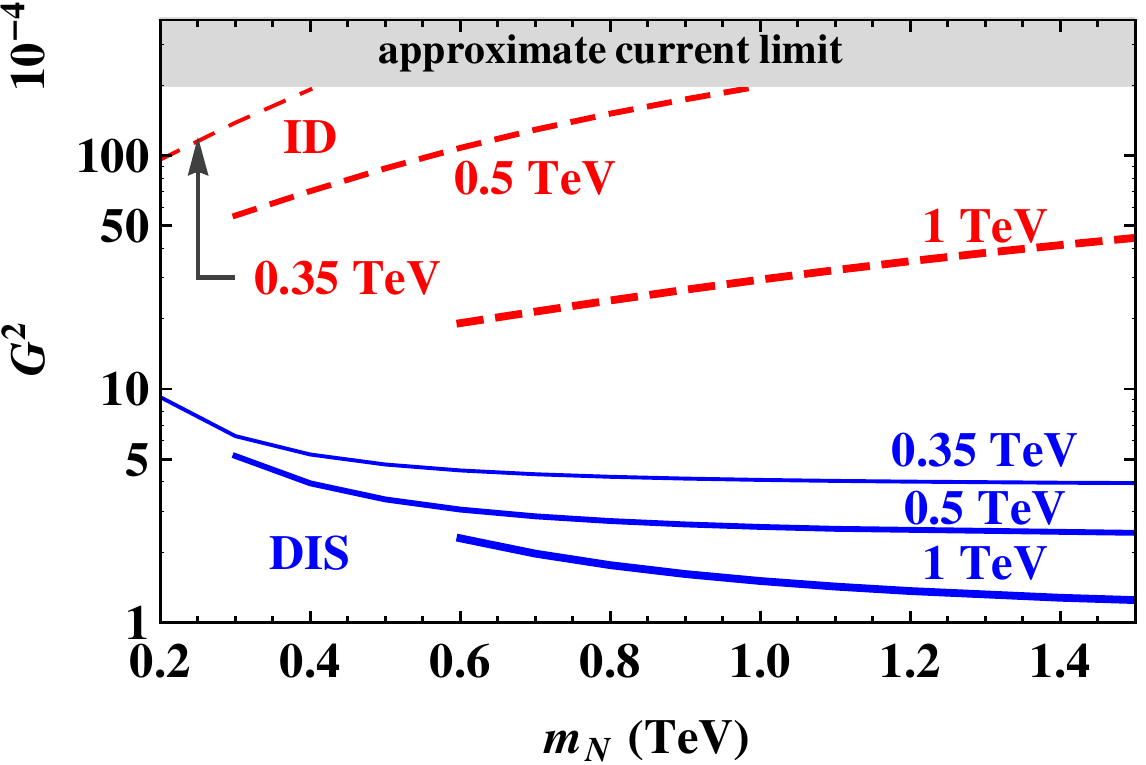}
}
\caption{\label{Fig:low-energy} Discovery (DIS) and identification (ID) reach on $G^2\equiv(G_L^{Ne})^2=(G_R^{Ne})^2$. The low-energy case (350~GeV) is compared with the nominal energy cases of 500~GeV and 1~TeV, all at an assumed integrated luminosity of $500~\text{fb}^{-1}$. The approximate current limit on these couplings is indicated as a grey band.}
\end{figure}

\section{Discovery and identification reach at $\sqrt{s}=350$~GeV}
\label{sect:low-energy}

In view of the possibility of a staged ILC construction, we would like
to comment on the possibility of obtaining bounds on heavy neutral
leptons at 350~GeV. As illustrated in Fig.~\ref{Fig:low-energy},
polarized beams would already at this low energy allow to place a limit
on possible $NWe$ couplings, in particular at low masses $m_N$.
In this figure we explore masses beyond the corresponding
kinematical reach.  Even at this rather low energy there is already sensitivity to discover
heavy lepton couplings
in the range of $G^2\sim 10^{-3}$ for low masses and up to $G^2\sim 5 \times 
10^{-4}$ for heavy masses $m_N$ and with an assumed integrated luminosity of 500 fb$^{-1}$. 
It is seen that one can identify heavy-lepton-mixing effects for masses up to $m_N\sim400~\text{GeV}$.

Discovery is seen to become more sensitive at higher masses, since the effect is approximately proportional to $1-r_N$, whereas for identification the sensitivity is governed by $r_N$, and thus becomes less efficient at higher masses.
For higher beam energy, both sensitivities improve.
\section{Concluding remarks}
\label{sect:conclusions}

In this note we have studied the process $e^+e^-\to W^+W^-$ and seen
how to uniquely identify the indirect (propagator and exotic-lepton
mixing) effects of a heavy neutral lepton exchange in the $t$-channel.
Discovery of new physics, meaning exclusion of the Standard Model,
does not depend on having both initial beams polarized, but the sensitivity is improved with beam polarization.
Such ``discovery'' could be due to the existence of a $Z^\prime$, anomalous
gauge couplings, or the effect of a heavy neutral lepton.  The
potential of the ILC  to discover heavy lepton effects depends on the possible presence of a $Z^\prime$ contribution, and is vastly improved if one is able to
determine the polarization of the produced $W$s.

Identification of such new physics effect as being due to a heavy
neutral lepton exchange, as opposed to a $Z^\prime$ or AGC can be
achieved via the determination of a double polarization asymmetry.
This identification of heavy-lepton admixture is independent of the strength of any $Z$-$Z^\prime$ mixing, as well as the $Z^\prime$ model, but
requires having both initial beams polarized.

\section*{Acknowledgements}
\par\noindent
 This research has been partially supported by the Abdus Salam
ICTP under the TRIL and Associates Scheme and the Belarusian
Republican Foundation for Fundamental Research. The work of AAP
has been partially supported by the Collaborative Research Center
SFB676/1-2006 of the DFG at the Department of Physics, University
of Hamburg. The work of PO has been supported by the Research
Council of Norway.



\end{document}